\documentclass[twocolumn,pre,aps,tightenlines,floatfix,superscriptaddress,showpacs]{revtex4-1}
\usepackage{amsmath}
\usepackage{graphicx}
\usepackage{amssymb}
\usepackage{color,soul}
\usepackage[T1]{fontenc}
\usepackage{ae,aecompl}
\usepackage{hyperref}

\begin{document}

\title{Switching synchronization in 1-D memristive networks: An exact solution}

\author{V. A. Slipko}
\affiliation{Institute of Physics, Opole University, Opole 45-052, Poland}
\affiliation{Department of Physics and Technology, V. N. Karazin Kharkov National University, Kharkov 61022, Ukraine}
\author{Y. V. Pershin}
\email{pershin@physics.sc.edu}
\affiliation{Department of Physics and Astronomy, University of South Carolina, Columbia, South Carolina 29208, USA}
 \affiliation{Nikolaev Institute of Inorganic Chemistry SB RAS, Novosibirsk 630090, Russia}

\begin{abstract}
We study a switching synchronization phenomenon taking place in one-dimensional memristive networks when the memristors switch from the high to low resistance state.
It is assumed that the distributions of threshold voltages and switching rates of memristors are arbitrary. Using the Laplace transform, a set of non-linear equations describing the memristors dynamics is solved exactly, without any approximations. The time dependencies of memristances are found and it is shown that the voltage falls across memristors are proportional to their threshold voltages. A compact expression for the network switching time is derived.
\end{abstract}

\pacs{64.60.aq, 73.50.Fq, 73.63.-b, 84.35.+i}

\maketitle

\section{Introduction}

The collective effects in networks of resistors with memory (memristors~\cite{chua76a})
have attracted significant attention in recent years ~\cite{pershin11d,pershin13a,pershin13b,vourkas2014generalization,slipko15a,Caravelli17a,Slipko17a}  driven by their possible applications.
In particular, the most commonly studied networks of two nonvolatile memristors exhibit an amazing functionality of universal boolean logic~\cite{borghetti10a,kvatinsky2014memristor}.
Some volatile memristors~\cite{pershin17a} offer an alternative (but less practical) approach for the same application. It has been demonstrated theoretically that 2D memristive networks can be used to solve the shortest path~\cite{pershin13b} and maze~\cite{pershin11d} problems in one single step.
Moreover, it has been found that the simplest 1D memristive networks exhibit a complex switching dynamics~\cite{pershin13a} involving a switching synchronization phenomenon~\cite{slipko15a}, and 1D networks combining memristors and resistors can transfer and process information~\cite{Slipko17a}. Some of the recent advances in the area of memristor-based nanoelectronic networks are summarized in Refs.~\cite{adamatzky2013memristor,vourkas2016memristor}.

\begin{figure}[b]
\begin{center}
\includegraphics[width=.85\columnwidth]{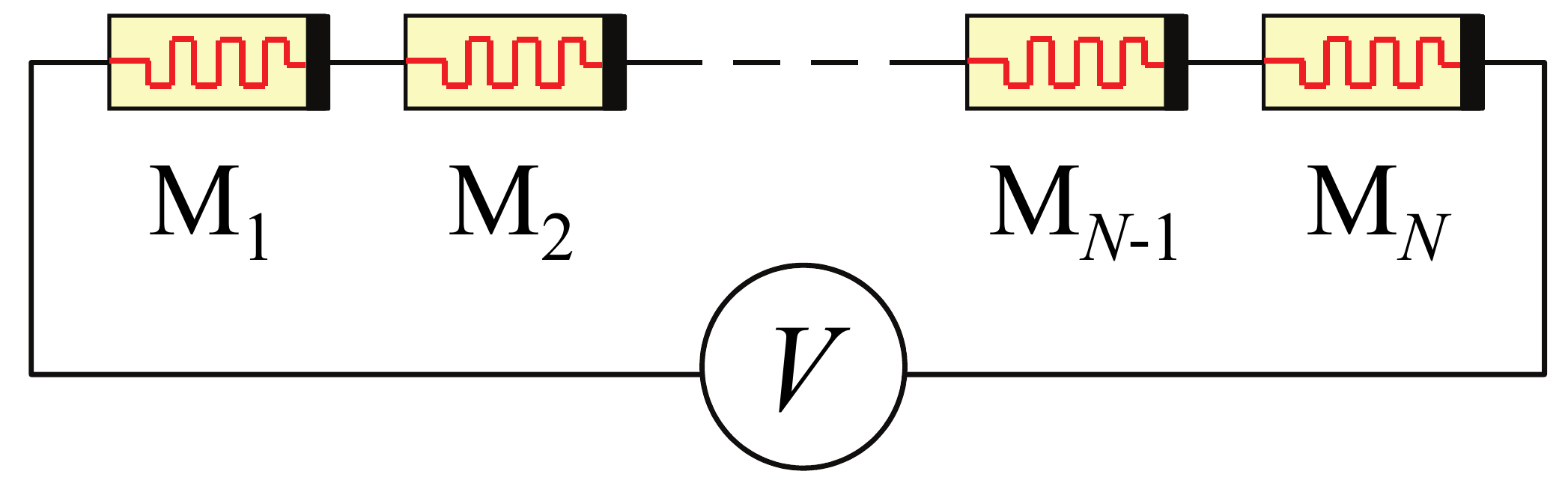}
\caption{\label{fig1} (Color online) One-dimensional network of $N$ memristive systems M$_i$ connected with the same polarity to the voltage source $V$. It is assumed that at the initial moment of time the memristors are in their high resistance states.}
\end{center}
\end{figure}

The switching synchronization~\cite{slipko15a} is a collective phenomenon taking place in 1D memristive networks (Fig. \ref{fig1}) in the regime when the threshold-type memristors with unequal switching rates switch from the high to low resistance states. It was shown in Ref.~\cite{slipko15a} that the switching of such memristors is synchronized such that the faster switching memristors 'wait' for the switching of slower ones. The details of switching synchronization phenomenon can be found in Ref. \cite{slipko15a} that studies a network of memristors with unequal switching rates but the same threshold voltages. The corresponding analytical results were derived in the limit when the applied voltage per memristor just slightly exceeds the memristor threshold voltage.

The goal of the present paper is to extend our previous result (Ref. \onlinecite{slipko15a}) to the case of memristors with unequal threshold voltages, unequal switching rates, and an arbitrary voltage applied to the network. In what follows, the equations describing the network dynamics are solved exactly using the Laplace transform. Based on this approach, we have been able to  find the time dependencies of memristances  exactly. Moreover, a generalized synchronization rule has been formulated. The main result of this paper is given by Eqs. (\ref{RiqFinal})-(\ref{DenomF}) that represent the time dependencies of memristances in a parametric form.

The voltage-controlled memristive systems are a class of two-terminal devices with memory defined by~\cite{chua76a}
\begin{eqnarray}
I&=&R^{-1}\left( x, V, t\right) V, \label{eq3} \\
\dot{x}&=&f\left(x, V, t\right), \label{eq4}
\end{eqnarray}
where $I$ and $V$ are the current through and voltage across the system, respectively, $R\left( x, V, t\right)$ is the memristance, $x$ is an $n$-component vector of internal state variables and $f\left(x, V, t\right)$ is a vector-function. For our purposes, it is sufficient to use the following simple model of memristors that, however, incorporates several important aspects of the physics of memristive devices such as the threshold-type switching, limited states of resistance, and finite switching time~\cite{pershin11a,di2013physical}. For $i$-th memristor, this model is formulated as~\cite{pershin09b}
\begin{eqnarray}
I_i&=&R_i^{-1}V_i
\label{eq:model1} \\
\frac{\textnormal{d}R_i}{\textnormal{d}t}&=&
\begin{cases} \pm\textnormal{sgn}(V_i)\beta_i(|V_i|-V_{t,i}) \;\textnormal{if} \;\; |V_i|>V_{t,i} \\ 0 \;\;\;\;\;\;\;\;\;\;\;\; \textnormal{otherwise} \end{cases} ,
\label{eq:model2}
\end{eqnarray}
where $I_i$ and $V_i$ are the current through and voltage
across $i$-th memristor, the memristance $R_i$ is used as the internal state variable \cite{chua76a}, $\beta_i$ is the switching rate, $V_{t,i}$ is the (positive) threshold voltage, $\textnormal{sgn}(.)$ is the sign of the argument, and $+$ or $-$ sign is selected according to the device connection polarity. Additionally, it is assumed that the memristance is limited to the interval [$R_{on}$, $R_{off}$], where $R_{on}$ and $R_{off}$ are the low and high resistance states of memristors, respectively.

\section{The model} \label{sec2}

\subsection{Equations}

We start from the system of $N$ nonlinear equations describing the evolution of Fig. \ref{fig1} memristors that
are coupled through the current (or, equivalently, the total resistance $R$):
\begin{eqnarray}
\dot{R}_i(t)&=&-\beta_i\left[ V \frac{R_i(t)}{R(t)}-V_{t,i}\right],~i=1,...,N,
\label{Ri(t)def}\\
R&=&\sum_{i=1}^{N}R_i.
\label{Rdef}
\end{eqnarray}
Moreover, it is assumed that at the initial moment of time $t=0$, all voltage drops
$V_i$ across memristors are larger than the corresponding (positive)
threshold voltages $V_{t,i}$, namely, $V_i(t=0)> V_{t,i}$. In this case,
one can show that also $V_i(t)> V_{t,i}$. Indeed, if at some moment of time  $V_i(t)=V_{t,i}$ for some $i$, then this particular
 memristance $R_i$ does not change during the subsequent infinitesimal time interval (see Eq. (\ref{Ri(t)def})).
However, the ratio $R_i(t)/R(t)$ can only increase resulting in $V_i>V_{t,i}$ in
the next moment of time.
In order to realize this regime of operation,  the applied voltage $V$ should exceed the total
threshold voltage, $V>V_{t}=\sum_i V_{t,i}$.

It is convenient to introduce a new independent variable $q$ instead of $t$,
$q=\int_0^t V \textnormal{d}t'/R(t')$, which represents the charge flown through the network by the time $t$.
Then, the system of Eqs. (\ref{Ri(t)def})-(\ref{Rdef}) can be transformed into
\begin{eqnarray}
\frac{\textnormal{d}R_i}{\textnormal{d}q}&=&-\beta_i\left[R_i-\frac{V_{t,i}}{V}R\right],
\label{Ri(q)def} \\
t&=&\frac{1}{V}\int\limits_0^q \textnormal{d}q' R(q').
\label{t(q)def}
\end{eqnarray}
Importantly, one can notice from the above equations that this change of independent variable has linearized the system of Eqs. (\ref{Ri(t)def})-(\ref{Rdef}).

\subsection{Laplace transform solution}

Next, we introduce the Laplace transforms of $R_i$ and $R$ as
$F_i(p)=\int_0^\infty R_i(q)\exp(-pq)\textnormal{d}q$ and $F(p)=\int_0^\infty R(q)\exp(-pq)\textnormal{d}q$, respectively.  Applying the
Laplace transformation to Eq. (\ref{Ri(q)def}) yields
\begin{equation}
F_i(p)=\frac{R_i(0)}{p+\beta_i}+\frac{V_{t,i}\beta_i}{V}\frac{F(p)}{p+\beta_i},
\label{Fi}
\end{equation}
where $R_i(0)$ is the initial memristance of $i$-th memristor.
The sum of Eqs. (\ref{Fi}) for $i=1,...,N$ results in the following expression
for the Laplace transform of the total memristance:
\begin{equation}
F(p)=\frac{\sum\limits_{i=1}^{N}\frac{R_i(0)}{p+\beta_i}}
{1-\frac{1}{V}\sum\limits_{i=1}^{N}\frac{V_{t,i}\beta_i}{p+\beta_i}}.
\label{F}
\end{equation}
One can notice that $F(p)$ is a rational function that approaches zero at infinity
as
\begin{equation}
F(p)=\frac{1}{p}\sum_{i=1}^{N}R_i(0)+O\left(\frac{1}{p^2}\right),~ p\rightarrow\infty.
\label{Finfty}
\end{equation}
Therefore, we can safely apply the inverse Laplace transform to obtain the
total memristance $R$ as a function of charge $q$ flown through
the network. This results in
\begin{eqnarray}
R(q)=\sum_{k}
 \textnormal{Res}(F(p)e^{pq};p_k)\nonumber\\
 =-\textnormal{Res}(F(p)e^{pq};p=\infty),
\label{R(q)}
\end{eqnarray}
where $p_k$ are the singular points of the function (\ref{F}).

\begin{figure}[tb]
\begin{center}
\includegraphics[width=.8\columnwidth]{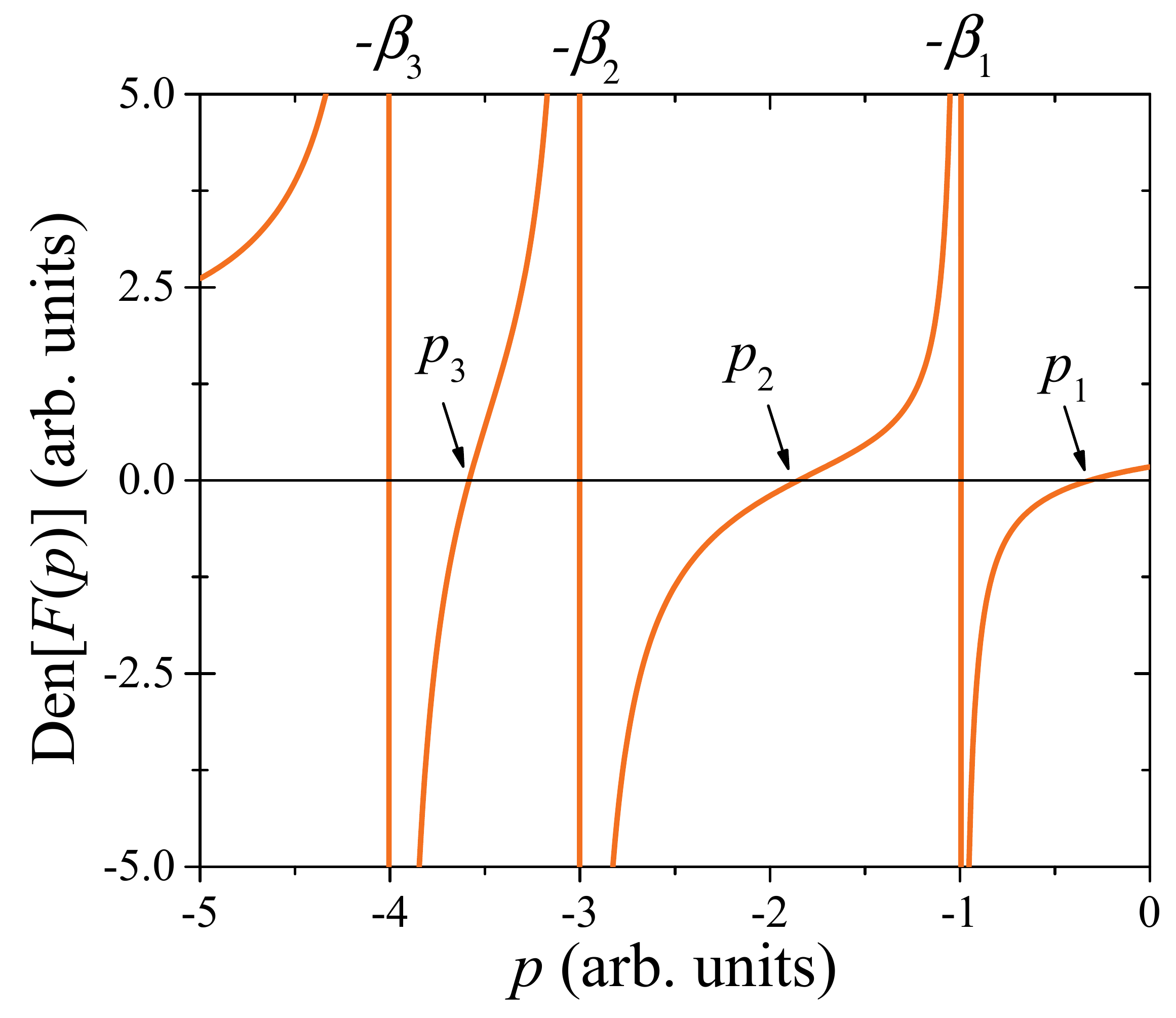}
\caption{\label{fig2} (Color online) Schematics of the denominator of $F(p)$  (Eq. (\ref{F})) for the case of $N=3$.}
\end{center}
\end{figure}

Then, by using Eq. (\ref{t(q)def}), one finds the time as a function of charge
\begin{eqnarray}
t(q)=\frac{1}{V}\sum_{k}
 \textnormal{Res}\left(\frac{F(p)}{p}(e^{pq}-1);p_k\right)\nonumber\\
 =-\textnormal{Res}\left(\frac{F(p)}{p}e^{pq};p=\infty\right).
\label{t(q)}
\end{eqnarray}
The conditions of positivity of the threshold voltages  $V_{t,i}$ and rates
$\beta_i$ guarantee that the points $p=-\beta_i$ are regular ones for the function
$F(p)$:
\begin{equation}
\lim\limits_ {p\rightarrow -\beta_i} F(p)=-\frac{R_i(0)V}{V_{t,i}\beta_i}.
\label{limbetai}
\end{equation}
So, the only singular points of the function (\ref{F})
are the zeroes of its denominator.
Clearly, there are no more zeroes than the number of different values among the rates $\beta_i$.
Moreover, all these zeroes are negative and they are separated by the corresponding
rates  $-\beta_i$. That can be clearly seen from the graph of the denominator
of Eq. (\ref{F})  as a function of $p$ (see Fig. \ref{fig2}). Let us assume that all $\beta_i$ are different.
In this case,
there are exactly $N$ different singular points $p_1, p_2, ..., p_N$, which
are simple poles satisfying the following inequalities:
\begin{eqnarray}
-\beta_N<p_N<-\beta_{N-1}<\cdots<-\beta_{1}<p_1<0 .
\label{ineq}
\end{eqnarray}

In order to find $R_i(q)$ we apply the inverse Laplace
 transform to $F_i(p)$ given by Eq. (\ref{Fi}) and obtain
\begin{eqnarray}
R_i(q)=\frac{V_{t,i}\beta_i}{V}\sum_{k}
 \textnormal{Res}\left(\frac{F(p)}{p+\beta_i}e^{pq};p_k\right)\nonumber\\
 =R_i(0)e^{-\beta_i q}-\frac{V_{t,i}\beta_i}{V}
 \textnormal{Res}\left(\frac{F(p)}{p+\beta_i}e^{pq};p=\infty\right).
\label{Riq}
\end{eqnarray}

Using the above mentioned information about the singular points of function
$F(p)$, the final results can be presented
as finite sums over the singular points of function
$F(p)$:
\begin{eqnarray}
R_i(q)&=&\frac{V_{t,i}\beta_i}{V}\sum_{k}
\frac{f_{k}}{p_k+\beta_i}e^{p_k q},
\label{RiqFinal}\\
t(q)&=&\frac{1}{V}\sum_{k}
 \frac{f_k}{p_k}e^{p_k q}+T_0,
\label{tq}\\
R(q)&=&\sum_{k}f_ke^{p_k q},
\label{Rq}
\end{eqnarray}
where the constants $f_k$ and $T_0$ are given by
\begin{eqnarray}
f_k&=&\textnormal{Res}(F(p);p_k)=\frac{V\sum_{i=1}^{N}\frac{R_i(0)}{p_k+\beta_i}}
{\sum_{i=1}^{N}\frac{V_{t,i}\beta_i}{(p_k+\beta_i)^2}},\\
\label{fk}
T_0&=&-\frac{1}{V}\sum_k \textnormal{Res}\left(\frac{F(p)}{p};p_k\right)\nonumber\\
&=&\frac{1}{V}\textnormal{Res}\left(\frac{F(p)}{p};p=0\right)
=\frac{\sum_{i=1}^{N}\frac{R_i(0)}{\beta_i}}{V-\sum_{i=1}^{N}V_{t,i}}.
\label{T0}
\end{eqnarray}
and $p_k$-s are the roots of the equation
\begin{eqnarray}
1-\frac{1}{V}\sum_{i=1}^{N}\frac{V_{t,i}\beta_i}{p+\beta_i}=0.
\label{DenomF}
\end{eqnarray}
The exact solution (\ref{RiqFinal})-(\ref{DenomF}) of the system of Eqs. (\ref{Ri(t)def})-(\ref{Rdef})  determines  the time dependencies of all memristances in the parametric form.

It is interesting to note that $T_0$ (given by Eq. (\ref{T0})) can be considered as a
characteristic switching time for the network, being the time when all resistances turn
to zero. In reality, however, the switching of any individual memristor stops when its resistance reaches the lowest possible value $R_{on}$.
Clearly, in the limit of $R_{on}\ll R_{off}$, the time $T_0$ provides a good estimate for the network switching time.

\subsection{Asymptotic behavior}

Next we consider the asymptotic behavior of the exact solution at long times. This is possibly the most
interesting case as the short time behavior is strongly influenced by the initial conditions.
Consider the case of $\beta_1 q\gg 1$. Then, as it follows from inequalities
(\ref{ineq}), all the terms in the sums (\ref{RiqFinal})-(\ref{Rq})
except for the first ones can be neglected as the contributions of the higher order terms
(corresponding to $p_i$ with $i\geq 2$) are exponentially small.
 Eliminating the charge $q$, we find the asymptotic behavior of the total resistance $R$  and the
individual memristances $R_i$:
\begin{eqnarray}
R=V|p_1|(T_0-t),
\label{R_large}\\
R_i=\frac{V_{t,i}}{V}\frac{\beta_i}{(\beta_i+p_1)}R.
\label{Ri_large}
\end{eqnarray}
It follows from Eq. (\ref{R_large}) that at long times
the total resistance of the network decreases linearly with time at the rate defined by $V |p_1|$. Moreover, Eq. (\ref{Ri_large})
demonstrates that the individual memristance $R_i$ evolves similarly to the total resistance $R$.

\begin{figure*}[tb]
\begin{center}
(a)\hspace{0.5cm}\includegraphics[width=.8\columnwidth]{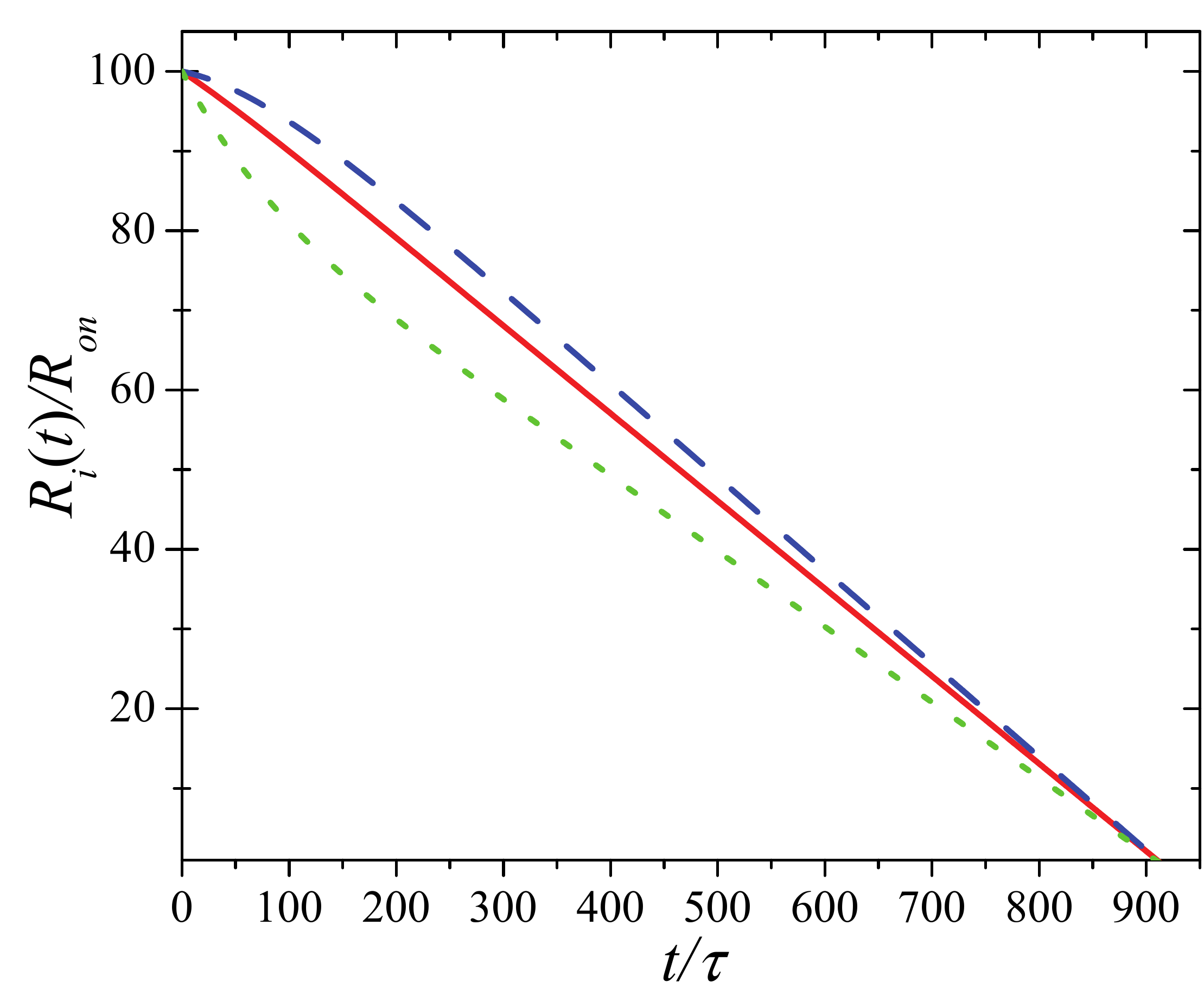}\;\;\;\;\;
(c)\hspace{0.5cm}\includegraphics[width=.8\columnwidth]{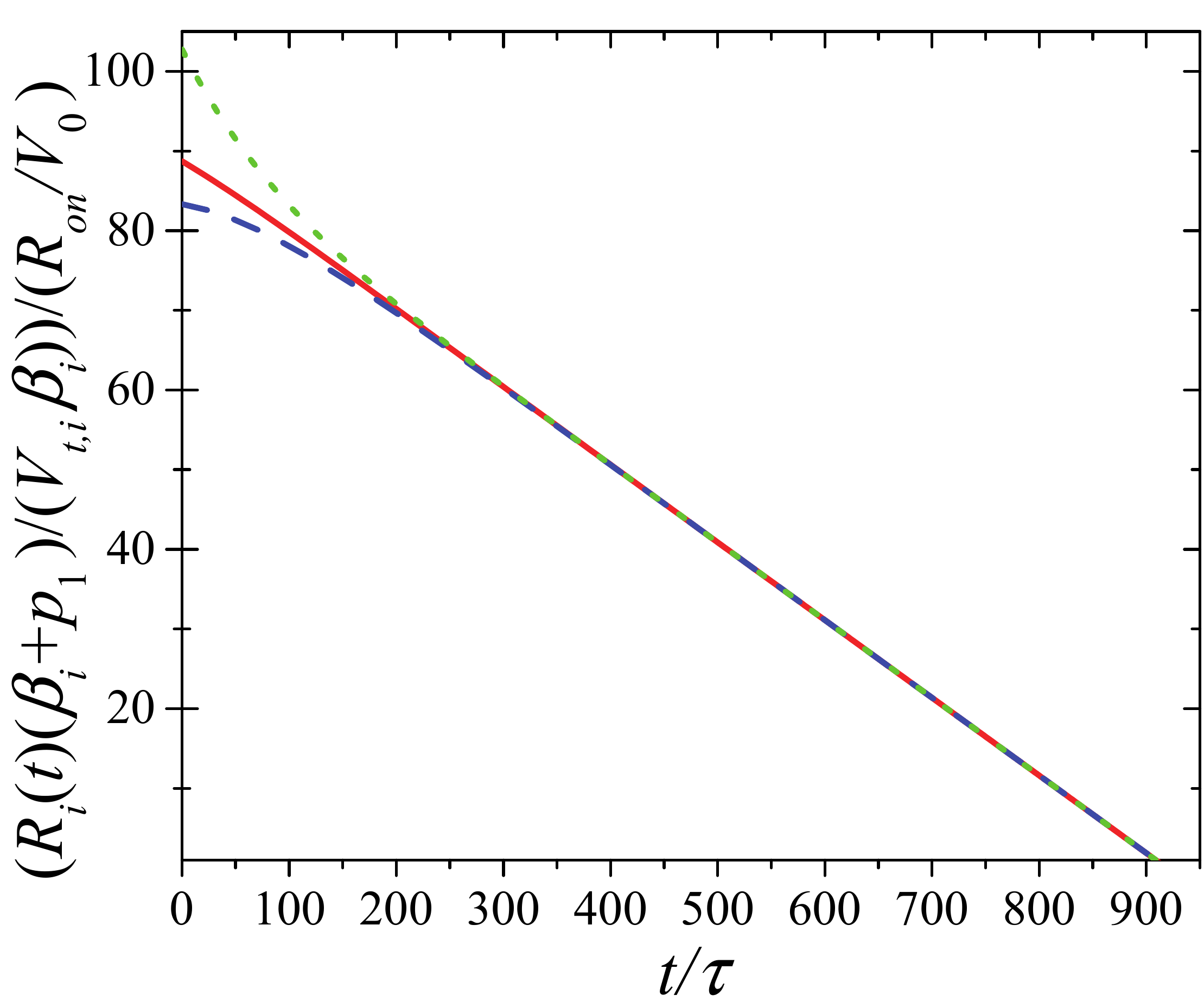}\\
(b)\hspace{0.5cm}\includegraphics[width=.8\columnwidth]{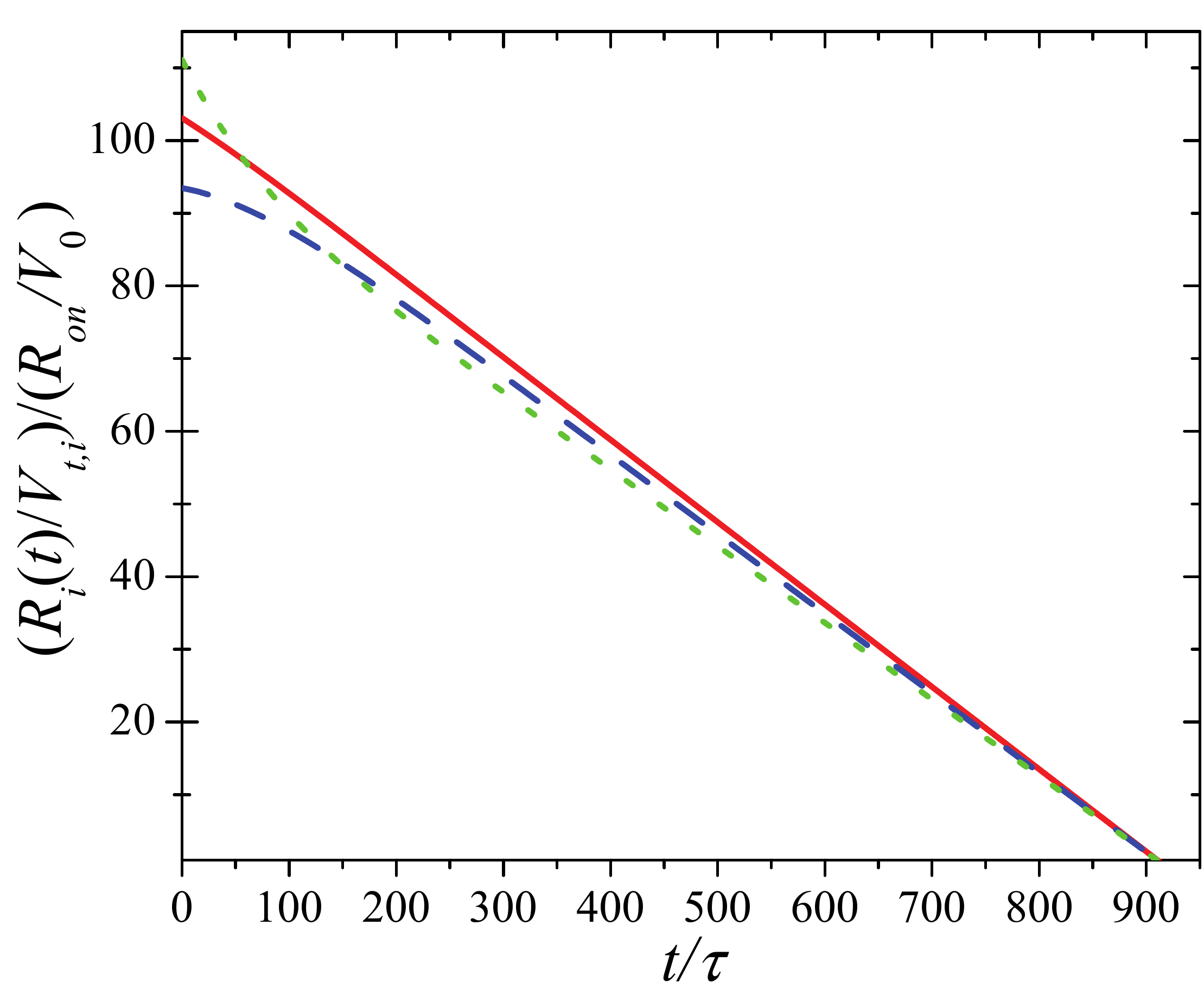}\;\;\;\;\;
(d)\hspace{0.5cm}\includegraphics[width=.8\columnwidth]{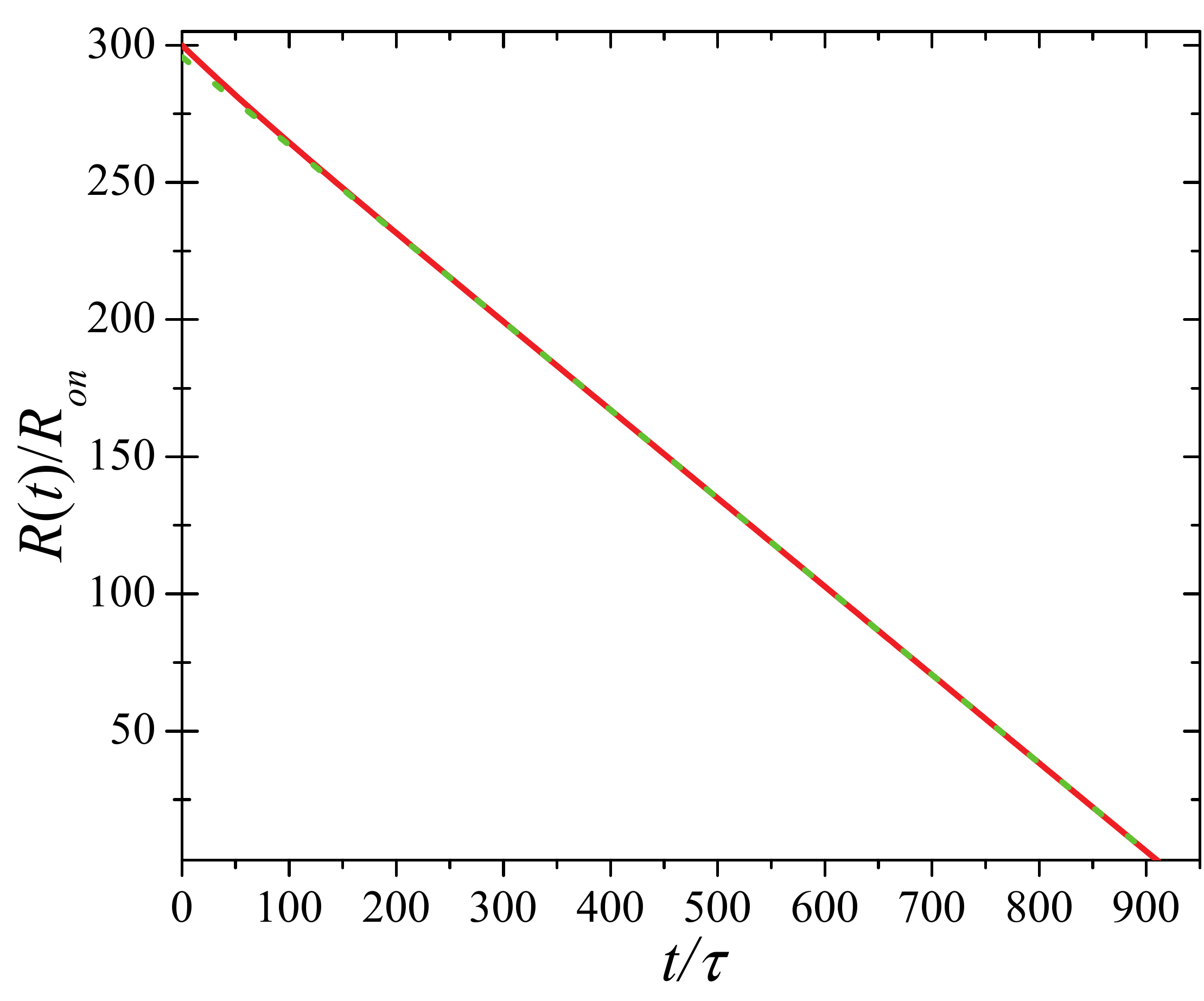}
\caption{\label{fig4} (Color online)
(a)-(c) Differently normalized $R_i(t)$ and (d) the total memristance $R(t)$ in $N=3$ network. See the text for details.}
\end{center}
\end{figure*}

Generally, the smallest root $|p_1|$ of Eq. (\ref{DenomF}) can not be found analytically.
Consider the special case of the applied voltage slightly exceeding the combined threshold voltage of the network,
namely, $\delta V=V-V_t\rightarrow +0$. In this case, it is possible to show that $|p_1|\rightarrow 0$.
Consequently, $|p_1|$ can be calculated by expanding  the left-hand side of Eq. (\ref{DenomF}) with respect to the small $p$:
\begin{eqnarray}
p_1=-\frac{\delta V}{\sum\limits_{i=1}^{N}\frac{V_{t,i}}{\beta_i}}
\left[1-
\frac{\sum\limits_{i=1}^{N}\frac{V_{t,i}}{\beta_i^2}}
{\left(\sum\limits_{i=1}^{N}\frac{V_{t,i}}{\beta_i}\right)^2}
\delta V\right]+O(\delta V^3),
\label{p1}
\end{eqnarray}
$\delta V\rightarrow +0$, where the second term in the brackets should be small in comparison with the first one, i.e. with $1$. In this limiting case Eq.
(\ref{Ri_large}) can be  simplified. Neglecting $|p_1|$ compared to $\beta_1$ we obtain
\begin{equation}
\frac{R_{i}}{V_{t,i}}=\frac{R}{V}.
\label{Ri_sinch}
\end{equation}
Thus in this case the ratio of resistivity $R_i$ of the individual memristor to its threshold voltage $V_{t,i}$ does not depend on the index $i$.
This observation can be considered as the generalized synchronization effect.

Figure \ref{fig4}(a)-(c) shows differently normalized memristances
in a sample $N=3$ network calculated by using Eqs. (\ref{RiqFinal})-(\ref{DenomF}).
In order to obtain this plot, the following set of parameter values was used:
$R_i(t=0)=R_{off}=100R_{on}$,
$V_{t,1}=0.97V_0$, $V_{t,2}=1.07V_0$, $V_{t,3}=0.9V_0$, $\beta_1=0.7 \beta_0$, $\beta_2=0.9 \beta_0$, $\beta_3=1.3 \beta_0$, $V=1.1NV_0$. From this figure we notice that while the individual memristances $R_i(t)$ (Fig. \ref{fig4}(a)) exhibit quite a different evolution, the normalization by threshold voltages
$V_{t,i}$ (Fig. \ref{fig4} (b)) or even better  the combination
$R_i(\beta_i+p_1)/(V_{t,i}\beta_i)$ (Fig. \ref{fig4}(c)) puts the curves very close to each other (after an initial time interval).

Fig. \ref{fig4}(d) presents the total memristance calculated with  Eq. (\ref{Rq}) (solid line) alongside with its asymptotic behavior (dotted line, Eq. (\ref{R_large})). Clearly, there is an excellent agreement between these two results especially at longer times. It is interesting to note that for the selected set of parameters, the smallest root $|p_1|$ of Eq. (\ref{DenomF}) is approximately equal to $-0.096\beta_0$. This value is small compared to the rate $\beta_1=0.7 \beta_0$. Therefore, one can also use the asymptotic  Eq. (\ref{p1}), which results in $|p_1|=-0.098\beta_0$, so that the relative error is less than $1.6$ percent.

\section{Discussion}

\begin{figure}[tb]
\begin{center}
(a)\hspace{0.5cm}\includegraphics[width=.8\columnwidth]{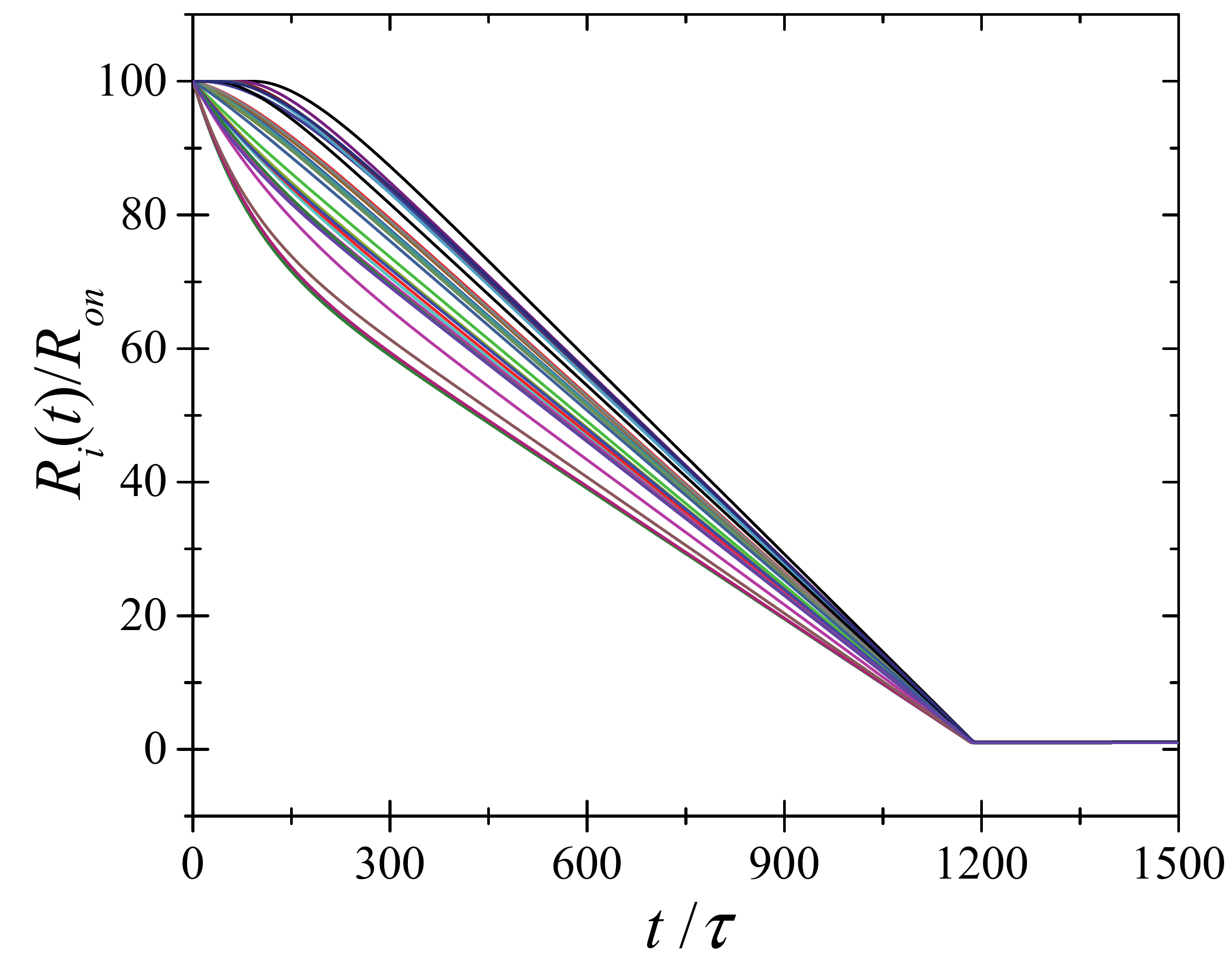} \\
(b)\hspace{0.5cm}\includegraphics[width=.8\columnwidth]{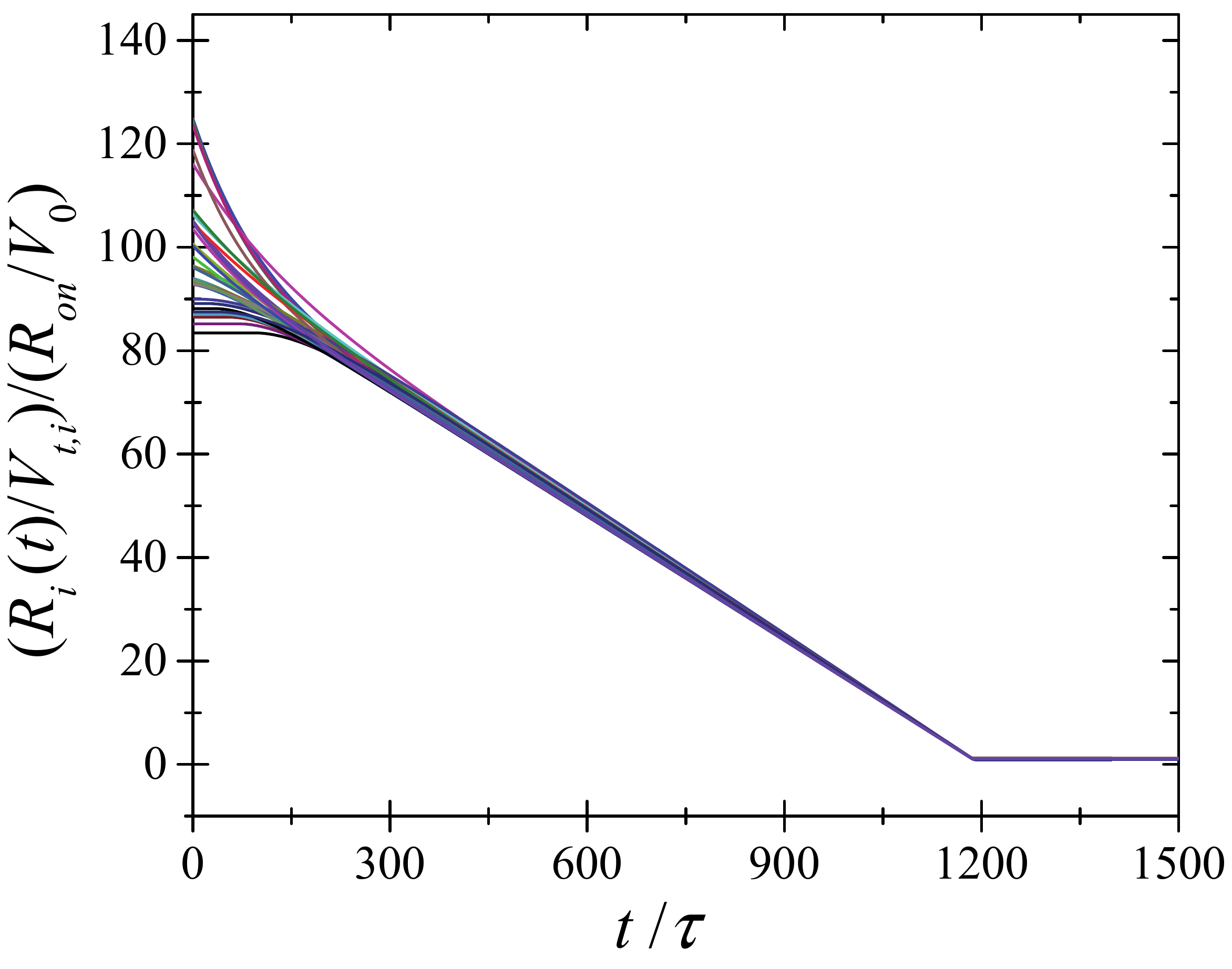}
\caption{\label{fig3} (Color online) Time dependencies of memristances (a) and normalized memristances (b) found in numerical simulations of the network dynamics. See the text for details.}
\end{center}
\end{figure}

The presented in Sec. \ref{sec2} exact solution of the non-linear problem of network dynamics (Eqs. (\ref{Ri(t)def})-(\ref{Rdef})) is given by Eqs. (\ref{RiqFinal})-(\ref{Rq}),
which are the main mathematical result of this paper. According to these equations, the time-dependencies of memristances are expressed in the parametric form through the charge flown through the network. This result is quite interesting by itself as $R_M(q)$ dependence is the main characteristic of the ideal memristor model~\cite{chua71a}.
While, in general, the memristors described by Eqs. (\ref{eq:model1})-(\ref{eq:model2}) are not ideal, their collective dynamics (in the situation considered in this work)  keeps them operating in the ideal memristor mode during the network switching process.

Moreover, we have identified the generalized switching synchronization condition (Eq. (\ref{Ri_sinch})). This condition extends the result of Ref. \cite{slipko15a} for the case of distributions of threshold voltages. Eq. (\ref{Ri_sinch}) shows that the ratio of the memristance to threshold voltage is the same for all memristors in the network at any moment of time (at longer times). In other words, the voltage across any memristor stabilizes in the proximity of its threshold voltage and stays at this value until the switching ends. Such a behavior can be explained by the collective feedback of network.

In order to demonstrate this feature graphically for a larger memristive network, we have performed numerical simulations of a network of $N=30$ memristors assuming flat distributions of switching rates and threshold voltages. According to Fig. \ref{fig3}(a), the switching synchronization can not be recognized in the dynamics of memristances, which evolve differently. The generalized switching synchronization phenomenon is clearly visible in Fig. \ref{fig3}(b) presenting the memristances normalized by threshold voltages. Starting at $t\sim 200 \tau$, where $\tau=R_{on}/(\beta_0V_0)$, the normalized memristances decrease coherently at the same rate.  Fig. \ref{fig3} was obtained with the threshold voltages and switching rates of memristors selected (with uniform distributions) in the intervals $[V_0-\delta V,V_0+\delta V]$ and $[\beta_0-\delta \beta,\beta_0+\delta \beta]$, respectively.
The following set of parameter values was used: $R_{off}=100R_{on}$, $\delta V=0.2V_0$, $\delta \beta=0.3 \beta_0$, $V=1.1NV_0$.
Clearly, the results presented in Fig. \ref{fig3} confirm our analytical predictions.

\section{Conclusion}
In conclusion, we have considered a simple  1D memristive network that, however, exhibit a very interesting switching synchronization  phenomenon. Using a suitable change of variable and the theoretical method of Laplace transform, we were able to solve a complex nonlinear problem analytically, what is surprising by itself, as the exact analytical results are known for a quite few number of nonlinear problems. Our analytical results are in agreement with numerical simulations.

\section*{Acknowledgment}
This work has been partially supported by the Russian Science Foundation grant No. 15-13-20021.

\bibliographystyle{apsrev4-1}
\bibliography{memcapacitor}

\end{document}